\documentclass[a4paper]{aipproc}

\usepackage{amsmath}
\usepackage{epsfig}
\usepackage{psfrag}
\layoutstyle{6x9}
\newcommand{\bm}[1]{\mbox{\boldmath $#1$}}

\begin{document}

\title{The Uniqueness of the $\Theta^+$ 
Pentaquark\footnote{
Invited talk at the 19. European Conference on Few-Body Problems in
Physics, Groningen (The Netherlands), 23-27 August 2004}
}

\author{P.J. Mulders}{
  address={Department of Physics and Astronomy, Vrije Universiteit Amsterdam,\\
De Boelelaan 1081, 1081 HV Amsterdam, The Netherlands}
}

\author{P. Jim\'enez Delgado}{
  address={Department of Physics and Astronomy, Vrije Universiteit Amsterdam,\\
De Boelelaan 1081, 1081 HV Amsterdam, The Netherlands}
}
                                                                                
\begin{abstract}
The existence of the $\Theta^+$ pentaquark
requires a peculiar mechanism to explain its stability. Looking at quark
clusters, notably diquark and triquark configurations, such a mechanism
may be found in the color-magnetic interaction between quarks.
It is possible to understand why the $\Theta^+$ is unique. Chiral dynamics,
in particular the ease of pion emission, will render other members of
the same flavor antidecuplet, such as the $\Xi^{--}$ very unstable.
\end{abstract}
                                                                                
\maketitle
                                                                                
\section{Introduction}\label{introduction}

The existence of a $\Theta^+(1540)$ baryon with $I=0$ and $Y=2$ 
decaying into a kaon and a nucleon with a small width of less than 10 MeV,
has been confirmed by several experiments \cite{NakanoLEPS,
BarminDIANA,StepanyanCLAS,BarthSAPHIR,Asratyan,KubarovskyCLAS,
AirapetianHERMES,AleevSVD}. 
A problem is the distribution of masses, which spans the region from 
1520 to 1540 MeV, but not in a way consistent with one object \cite{Karliner}.

Baryon and meson resonances in hadron physics generally appear 
in SU(3) flavor multiplets. The lowest-dimensional possibility for the
$\Theta^+(1540)$ is a flavor antidecuplet ($10^\ast$). Weak evidence
exists for another exotic member in such an antidecuplet in the
form of a $\Xi^{--}(1862)$ resonance \cite{AltNA49}, although 
there are by now many experiments that don't see this state.

Interesting fact about the $\Theta^+(1540)$ was the accurate
prediction made in the chiral soliton model \cite{Diakonov:1997mm}, including
the prediction of weak coupling to the $KN$-channel. The presence
of anti-strangeness in the system, however, calls for an understanding
in terms of quark models.

Baryons with $Y=2$ require a minimal configuration containing
four quarks and one antiquark. This opens up a multitude of
possibilities and it is therefore surprising that there does not
seem to be a rich spectrum of pentaquark states.
The general qualitative understanding of the absence of multi-quark
states (color singlet combinations other than $q\overline q$ and
$qqq$ or $q^3$) is the possibility of fission of such configurations
into simpler color singlets, $q^4\overline q \rightarrow (q^3)(q\overline q)$
or $q^6 \rightarrow (q^3)(q^3)$. For such a mechanism one naturally
expects widths larger than those for decays of baryons or mesons
via $q\overline q$-creation, thus several hundreds MeV, unless
the state is less than a few tens of MeV above threshold.

Actually, multiquark configurations may not be visible at all.
The possibility of fission spoils the confinement, which is the
necessary ingredient to produce hadrons. A way to deal with 
the 'artificial' confinement in the case of multiquark states
is the P-matrix formalism \cite{JaffeLow,Bakker} 
in which the 'calculated' configurations appear as the poles 
in the P-matrix. This quantity merely describes the boundary condition
needed to match a short-range quark description onto hadronic 
decay channels and the poles do not (necessarily) show up as
poles in the (physical) S-matrix.

\begin{figure}
  \includegraphics[width=7cm]{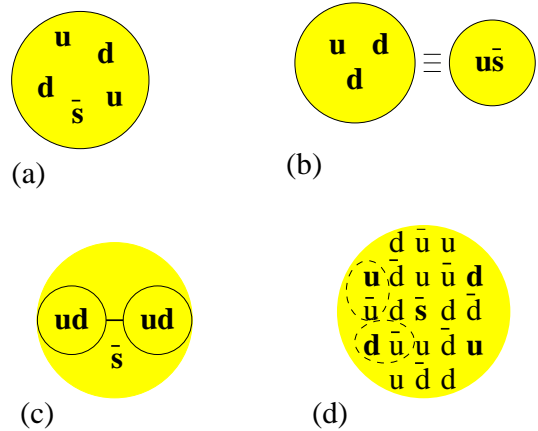}
  \caption{\label{structure}
   Possible structures of pentaquark baryons}
\end{figure}
                                                                                
Of course, the mechanism of fission is also impossible
if the mass is below the threshold of the fission products.
An example is the deuteron containing six quarks, which is stable
against decay into two nucleons. The example of the deuteron also
shows that a multi-quark configuration still allows many classes of
configurations (see Fig.~\ref{structure})
including bound states of color singlet subsystems.
This has been studied for the $KN$ system already in the sixties
(see e.g.~\cite{Dullemond})
The small width of the $\Theta^+(1540)$ points to a specific
internal structure that evades the fission procedure, such as was
for instance suggested in the work of Jaffe and Wilczek \cite{Jaffe:2003sg}.
This still asks for an explanation of the absence of rich
spectra of multi-quark states such as studied already decades ago
\cite{Jaffe:1976ih,Mulders:1979ea}.

Two possible ingredients that are relevant in a qualitative
analysis of the stability of multi-quark states, are
\\
(1) The first is the observation that hadron spectra follow a
constituent quark behavior with superimposed on that a color-magnetic
spin-spin interaction.
\\
(2) The second ingredient is the chiral symmetry implimentation in
the form of the light pions, of which the masses are lighter than 
expected on the basis of (1).
                                                                                
\section{Stability of multiquark configurations}

I will combine results of previous systematic studies of the spin-spin 
interaction in multi-quark clusters \cite{Jaffe:1976ih, Mulders:1979ea}
to look for exceptional configurations that experience a strong attractive
color-magnetic interaction and that cannot be lowered by fissioning the
system, adding (anti-)quarks, or emitting pions.
The hyperfine structure in the spectrum of baryons and mesons can be
understood in terms of the color spin-spin interaction arising from 
lowest-order one gluon exchange between pairs of quarks \cite{DeRujula:1975ge}
\begin{equation}
\Delta E_{cm} = - \sum_{i>j} M_{ij}~ (F_i^aF_j^a)
(\bm \sigma_i\cdot\bm\sigma_j)
\label{finemul}
\end{equation}
where $\sigma_i$ are the spin matrices and $F_i^a$ are the color charges of 
the \emph{i}th constituent quark and $M_{ij}$ measures the interaction 
strength, which in a specific model can be calculated and depends then
e.g. on quark masses and/or size parameters.We will simply use an
average strength in our estimates, in which case the color-magnetic
interactions becomes proportional to a group-theoretical factor,
\begin{equation}
\Delta =
-\sum_{i>j}(F_i^aF_j^a)(\bm \sigma_i\cdot\bm\sigma_j).
\label{fine}
\end{equation}
This expression will also be used for colored quark-clusters in
which all quarks are assumed to be in relative s-waves.

\begin{figure}
  \includegraphics[width=5cm]{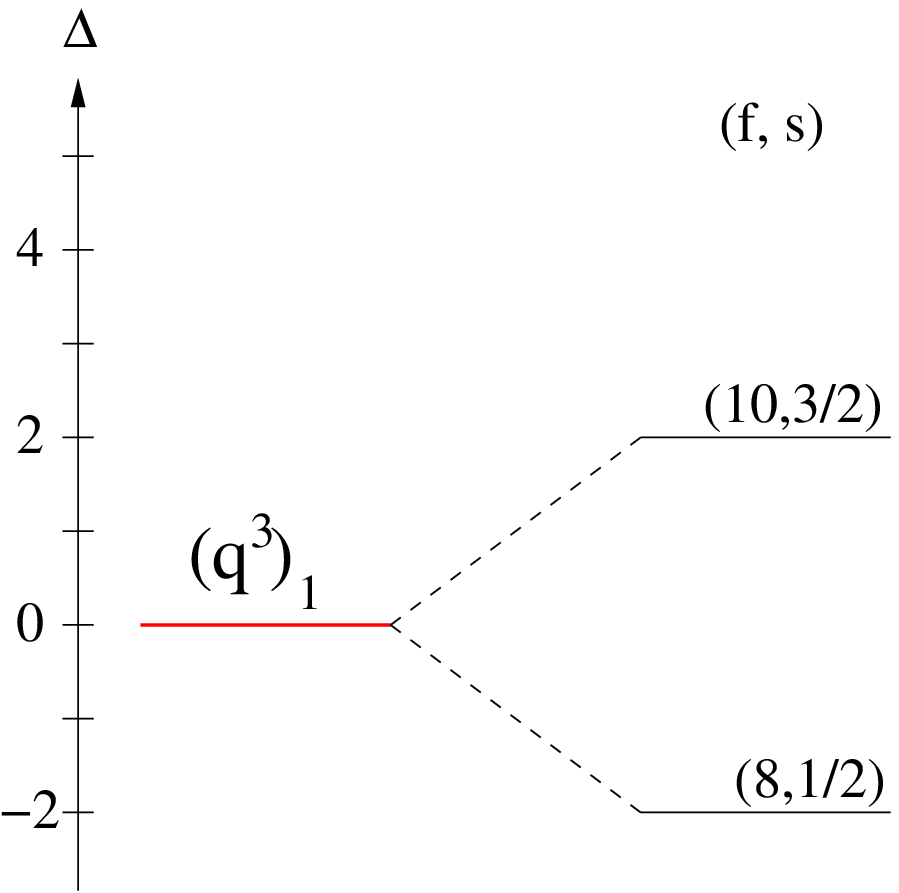}\hspace{1.5cm}
  \includegraphics[width=5cm]{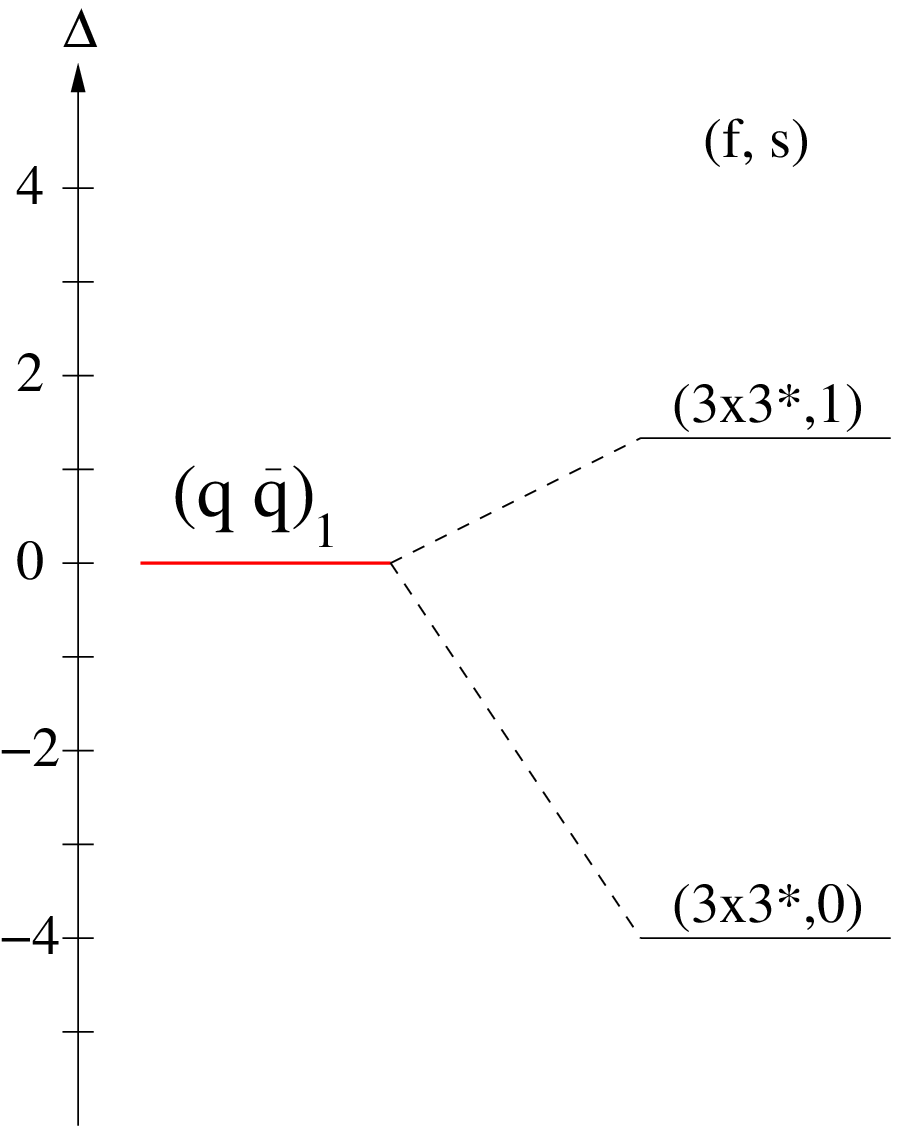}
  \caption{\label{fig-hadrons}
   Color magnetic hyperfine structure in baryons and mesons.
The configurations are labeled by flavor and spin (f,s)}
\end{figure}
                                                                                
The color-magnetic interaction in colorless
$(q\overline q)_{\underline 1}$ and
$(q^3)_{\underline 1}$ situations (mesons and baryons) is given in
Fig.~\ref{fig-hadrons}. It explains e.g. the splitting between
the baryon octet and decuplet, such as the $N-\Delta$ splitting of
about 300 MeV. Hence the proportionality constant relating the 
hyperfine structure and the factor $\Delta$ is about 75 MeV. This
also works fine for mesons like $K$ and $K^\ast$. With a flavor
independent strenght one should not expect extreme accuracy as
is clear from the fact that $\Lambda$ and $\Sigma$ baryons would
remain degenerate instead of the observed 70 MeV splitting, but
this can be cured by using flavor dependence, in particular a
slight decrease of the strength for strange quarks. At this stage
we will not worry about these details. A notable exceptional hadron
is the pion. The pion is much lighter (and $\pi - \rho$ splitting
is much larger) than expected on grounds of
the naive use of the color-magnetic interaction.

\begin{figure}
  \includegraphics[width=7cm]{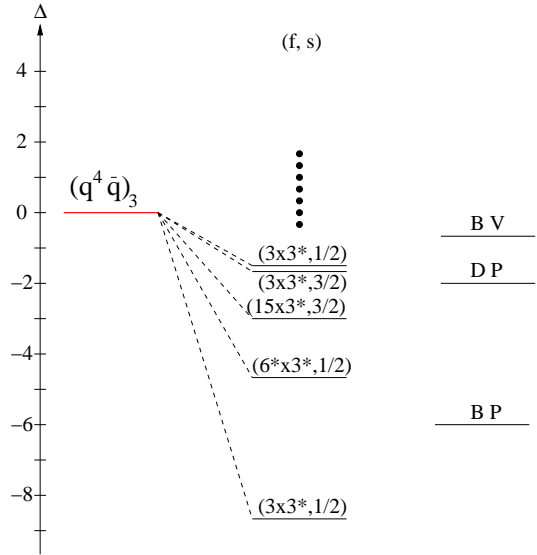}
  \caption{\label{fig-multiquarks}
   Color magnetic hyperfine structure for multiquark baryons.
The configurations are labeled by flavor and spin (f,s)}
\end{figure}
                                                                                
The calculation of color-magnetic hyperfine structure for s-wave
$(q^4\overline q)_{\underline 1}$ baryons was studied by Jaffe
in the bag model \cite{Jaffe:1976}.
The results for $\Delta$ are shown
in Fig.~\ref{fig-multiquarks} and show the lowest configuration 
with $\Delta = -9.045$. This, however, has flavor structure
$\underline{3}\times\underline{3}^\ast = \underline{1} + 
\underline{8}$, hence non-exotic. It has been suggested that
the $q^4\overline{q}$-configuration may be an important component of
the $\Lambda(1405)$ hyperon near the $\overline K N$-threshold.
The configuration with flavor 
$\underline{6}^\ast\times\underline{3}^\ast = \underline{8} + 
\underline{10}^\ast$ does contain a $\Theta^+$-candidate with
spin-parity $1/2^-$, but it is unstable against fission into
$KN$ (indicated as BP = Baryon-Pseudoscalar meson threshold in 
Fig.~\ref{fig-multiquarks}).

\begin{figure}
  \includegraphics[width=4.5cm]{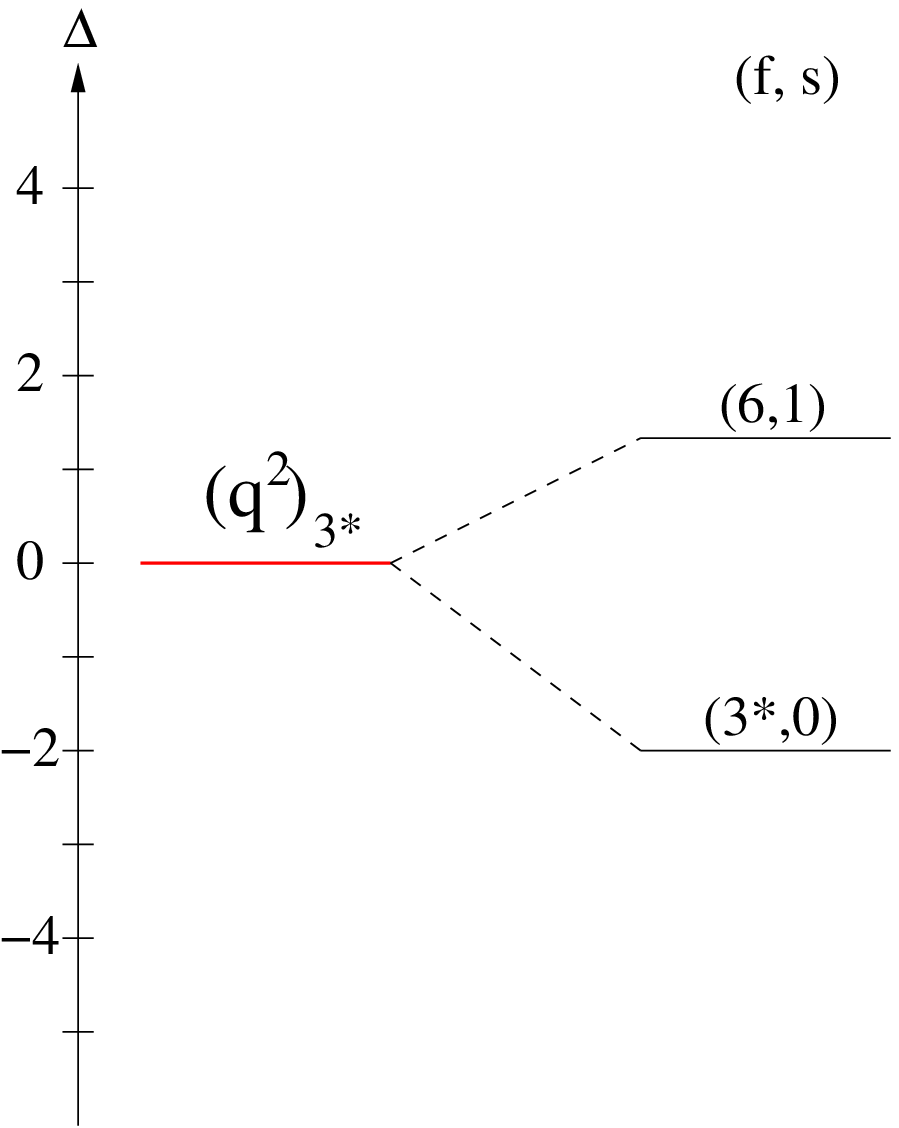}\hspace{1cm}
  \includegraphics[width=9.5cm]{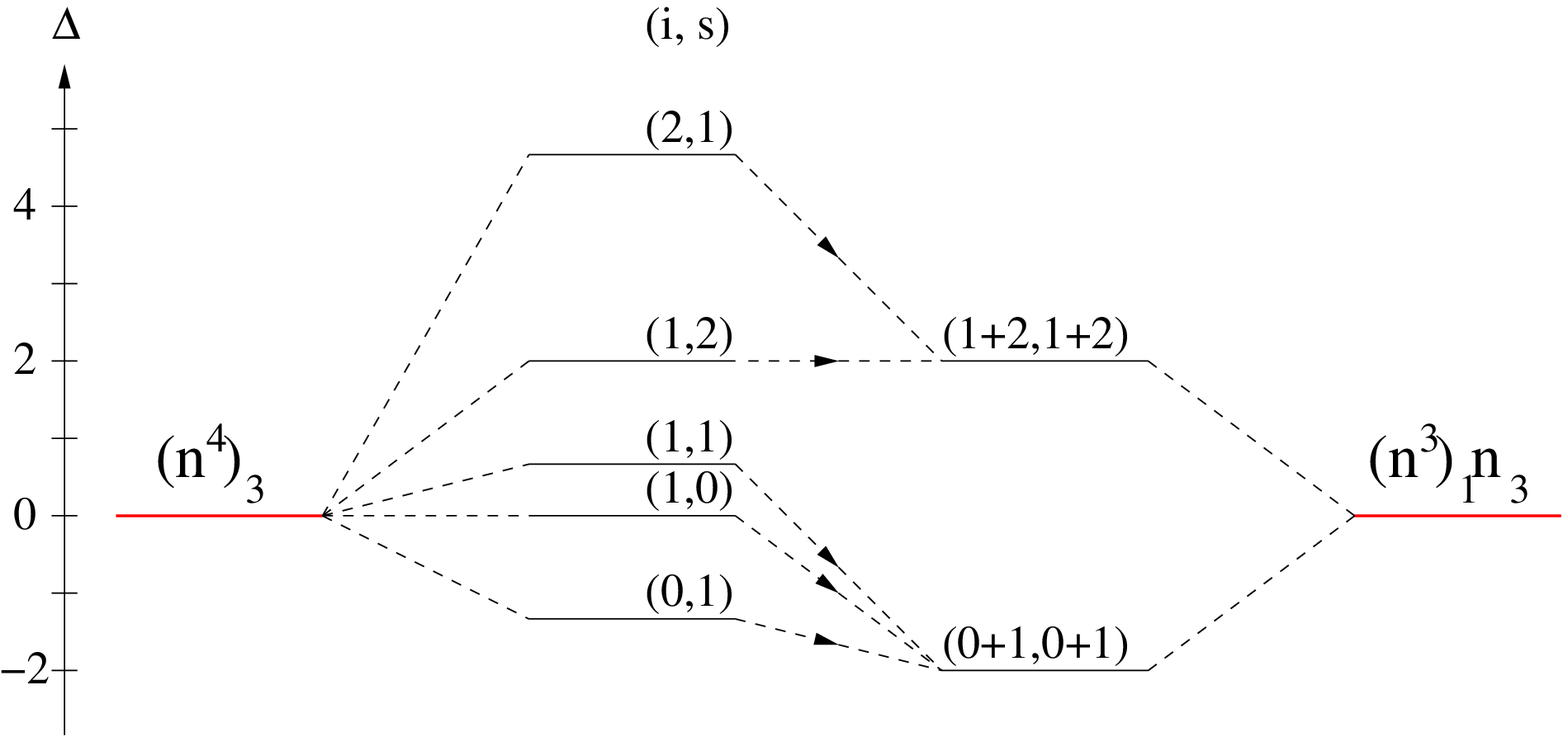}
  \caption{\label{cluster}
   Color magnetic hyperfine structure for two-quark and four-quark
configurations, the latter only for nonstrange quarks..
The configurations are labeled by flavor and spin (f,s) or
isospin and spin (i,s)}
\end{figure}
                                                                                
Turning attention to colored clusters, it is long known that the
color magnetic interaction for $(q^2)_{\underline 3^\ast}$ prefers
spin 0 diquarks (with antisymmetric flavor $\underline 3^\ast$ or 
for nonstrange quarks isospin 0) over spin 1 diquarks (Fig.~\ref{cluster}),
which has been the basis of the expectation of a special role for 
spin 0 over spin 1 diquarks in hadron spectroscopy or in color
superconductivity \cite{Jaffe:2003sg, Shuryak:2003zi}. 
Actually, having more quarks with a limited number of flavors,
in general, leads to color-magnetic repulsion. This phenomenon
is for instance visible for $(q^6)_{\underline{1}}$ (six quark) clusters
with for only nonstrange quarks, for which the lowest possibility is
$\Delta = + 2/3$, far above the $\Delta = -4$ of the (open) baryon-baryon
channel, which is the qualitative explanation of the repulsive core
in the nucleon-nucleon interaction at the quark level. Also for 
$(q^4)_{\underline 3}$ the lowest possibility ($\Delta = -4/3$) allows
fission $(q^4)_{\underline 3} \rightarrow
(q^3)_{\underline 1}\,(q)_{\underline 3}$ (see Fig.~\ref{cluster}).

\begin{figure}
  \includegraphics[width=9cm]{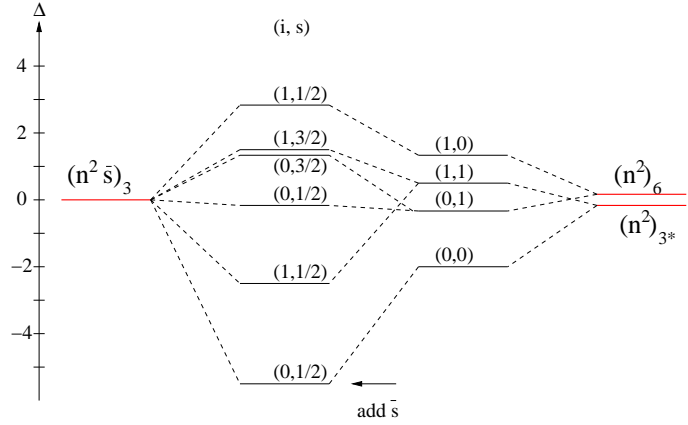}
  \caption{\label{level-4}
   Color magnetic hyperfine structure for $nn\overline s$-
configurations compared with the diquark structure.
The configurations are labeled by isospin and spin (i,s)}
\end{figure}
                                                                                
Looking at
$(q^2)_{\underline 3^\ast}\,(q^2)_{\underline 3^\ast}$ configurations
one of course can obtain a lower configuration with $\Delta = -4$.
For this one needs a spatial wave function that
reduces the overlap between the clusters, turning off the
short-range color-magnetic interaction between quarks belonging
to the different clusters. Given the quantum numbers of the 
bosonic diquarks, one needs in that case an antisymmetric wave function,
$[(q^2)_{\underline 3^\ast}\,(q^2)_{\underline 3^\ast}]_A$, realized
e.g. through an orbital angular momentum $\ell = 1$.

Actually, adding a strange antiquark to the nonstrange diquark
leads to a strong enhancement of the color-magnetic interaction
(from $\Delta = -2$ to $\Delta = -5.42$),
as shown in Fig.~\ref{level-4}. This has led to suggestions for
triquark-quark configurations \cite{Karliner:2003dt,Hogaasen}.
Actually, it is important to note that because of the identical
structure of the two diquarks involved, the strange quark will
oscillate between the diquark configurations, a mechanism proposed
already in Ref.~\cite{deSwart:1980}. The situation resembles
systems like the $H_2^+$-ion ($p-e-p$) or $^9Be$ ($\alpha - n - \alpha$)
\cite{private}.

\section{Conclusions}

Quark clustering may provide an answer to the exotic structure of
the $\Theta^+$ pentaquark. The particular combination of two
nonstrange diquarks and a strange antiquark may be a rather unique
one. For instance the $[ud]\overline s$-configuration is stable
against fission into $u + K^-$ or $d + K^0$ (with $\Delta = -4$). Although
flavor symmetry is usually a good guidance in hadron spectroscopy,
it may fail because the structure of $[ds]\overline u$ that would
be the one in the exotic $\Xi^{--}$ pentaquark would most likely
because of the exceptional role played by pions be very unstable
against fission into $s + \pi^-$. Hence the uniqueness could even
prevail over flavor symmetry. If established beyond any doubt
one shouldn't expect a proliferation of such states. 
The mechanism to understand the qualitative clustering picture
alluded to in this talk, however, would be a challenge in the study of
nonperturbative QCD.
\\[0.5cm]
We acknowledge discussions with Daniel Boer and Marek Karliner.

\end{document}